# Mechanisms of Seasonal - ENSO interaction

Eli Tziperman

Department of Environmental Sciences

The Weizmann Institute of Science

Rehovot 76100, Israel

Stephen E. Zebiak, Mark A. Cane

Lamont-Doherty Earth Observatory

Columbia University

Palisades, NY 10964, USA

# Abstract

The mechanisms of interaction between the seasonal cycle and ENSO are investigated using the Zebiak and Cane ENSO prediction model. The most dominant seasonal effect is found to be due to the wind divergence field, as determined by the seasonal motion of the ITCZ, through its effect on the atmospheric heating. The next order seasonal effects are due to the seasonality of the background SST and ocean upwelling velocity, and the corresponding mechanisms are analyzed. It is suggested that the seasonal forcing has a first order effect on ENSO's dynamics. Important aspects of the seasonal forcing may be included in idealized delayed oscillator ENSO models by making the model background seasonally shift from stable to unstable states.

# 1 Introduction

The apparent partial locking of El Niño-Southern Oscillations (ENSO) events to the seasonal cycle, as expressed in their tendency to peak towards the end of the calendar year, is perhaps one of ENSO's most distinctive characteristics (Rasmusson and Carpenter, 1982). This partial locking is a clear indication that the seasonal cycle in the equatorial Pacific ocean and atmosphere plays a major role in ENSO's dynamics. This role, however, has been somewhat neglected in many theories of the ENSO cycle, implying that the seasonal cycle may not be essential to ENSO's onset and termination (e.g. Suarez and Schopf, 1988; Graham and White, 1988; Battisti and Hirst, 1989; Neelin, 1991).

Recently, seasonal forcing has been suggested as a possible reason for ENSO's irregularity (Tziperman et al., 1994, Jin et al., 1994, Tziperman et al. 1995, Chang et al., 1994, Chang et al, 1995). ENSO is presented in these theories as a nonlinear oscillator forced by the seasonal cycle in the equatorial Pacific. The ENSO oscillator can enter into a nonlinear resonance with the seasonal forcing. Such a resonance is characterized by perfectly periodic behavior, in exact phase with the seasonal cycle. For sufficiently nonlinear dynamics, several such resonances may coexist, and then the ENSO oscillator, not being able to prefer a single such resonance, jumps irregularly between the different resonances creating the observed ENSO irregularity. These theories, however, leave many important questions unanswered. In particular, they do not deal with specific physical mechanisms by which the seasonal cycle interacts with the interannual ENSO variability.

Even before the possible role of the seasonal cycle in forcing ENSO's irregularity was suggested, there have been quite a few attempts to discuss the dynamics of seasonal-interannual interactions. Philander (1983) noted that a major seasonal influence must be the wandering of the Pacific Inter-tropical Convergence Zone (ITCZ) and its effect on the atmospheric heating and the coupled instability believed responsible for ENSO's onset. Hirst (1986) noted that the annually averaged basic state of the equatorial Pacific is too stable to support the onset of ENSO as a coupled ocean-atmosphere instability. Thus the importance of the seasonality of the background state in creating times during the year in which this state is unstable and ENSO can initiate through a coupled instability mechanism (Philander, 1983; Philander et al, 1984; Hirst, 1986). Zebiak and Cane (1987) have suggested that the seasonal changes of the background climatology may be viewed as a seasonal modulation of the coupling strength between the ocean and the atmosphere. This approach was followed by Cane et al (1990) and Munich et al (1991) where simple delay models were used with a coupling coefficient that varied seasonally. Battisti (1989) emphasized the seasonality in the oceanic upwelling and defined a "potential instability index" reflecting the influence of the upwelling seasonality on the background stability. Predictability studies of ENSO also indicated the crucial role of the seasonal cycle. Blumenthal (1991) and Xue et al (1994) showed that the stability of the background state of the equatorial Pacific is strongly seasonal, and that this has implications on the seasonality of ENSO predictability. Related seasonal effects were also investigated by Goswami and Shukla (1991), Battisti and Sarachik (1995), Latif et al. (1994) and Chen et al. (1995).

In the present work we wish to identify the specific physical mechanisms by which the seasonality of the background state in the Equatorial Pacific affects the interannual ENSO variability. We are in particular interested in the mechanisms responsible for ENSO's locking to the end of the calendar year. This objective is approached using numerical experiments with the Zebiak and Cane (1987, hence after ZC) ENSO prediction model. The ZC model is especially convenient for our purposes because the background seasonality is specified rather than simulated. This background can therefore be easily modified to examine its importance in ENSO's dynamics.

We begin in section 2 by presenting the "standard" model solution, and describing in detail how the five background fields specified as monthly climatologies in the ZC model enter the model equations and parameterizations. Then, in section 3, we describe a set of experiments in which the background fields are set to be seasonal one at a time, with the others set to their annual mean. By comparing the results to the standard run we identify the main (first order) seasonal effect in the model to be the seasonal evolution of the wind divergence field. The divergence field evolution, reflecting the seasonality in the ITCZ location, affects the model ENSO dynamics through the atmospheric heating parameterization. It turns out, however, that while the model ENSO events with the background divergence the only seasonal field are quite reasonable, they are not satisfactorily close to the standard model solution and to the observed ENSO characteristics. We therefore proceed in section 4 to identify second order seasonal effects by searching for a second seasonal field that strongly interacts with the interannual variability. We find that this field is the SST and identify the precise mechanism in which the seasonality in the SST field determines ENSO's locking to the seasonal cycle. Finally (section 5), the seasonality in the upwelling velocity also turns out to have an important role in influencing the events amplitude and frequency.

Based on our analysis it is suggested that the seasonality of the equatorial Pacific should not be ignored even in simplified ENSO models (such as delayed oscillator models). Such models can incorporate the important seasonal effects by using a seasonally varying background that is unstable and enables the coupled ocean-atmosphere instability mechanism during some months of the calendar year, and that is stable and does not allow this instability to develop in other months. We conclude in section 6.

## 2  The model and standard model solution

The version of the ZC model used here was described by Zebiak and Cane, (1987, section 4c); its ocean and atmosphere components have been carefully tuned to be "optimal" in terms of fitting to observed ENSO characteristics. Fig. 1a shows a 30 year time series of the ZC model NINO3 index (averaged sea surface temperature over the model's East Pacific 5°S–5°N and 90°W–150°W). The power spectrum for this time series, based on a 1024 year model NINO3 time series is given in Fig. 1b. A histogram of the number of ENSO events occurring in each month of the calendar year during a 1024 year model integration is shown in Fig. 1c. Finally, a histogram of the number of occurrences of a given separation between ENSO events is given in Fig. 1d. An event is defined for our purpose here as a maximum of the NINO3 index.

Only NINO3 maxima that are larger than a specified threshold are considered "events" in our analysis. This threshold is chosen to be 1/2 of the average of the positive segments of the NINO3 time series. We found that the use of a threshold value or the precise choice of threshold did not change any of the results presented below. It should be noted that the calculated locking of ENSO events to the annual cycle as shown for example in the histograms of Fig 1c may, in principle, depend on the way the time series is analyzed. The model NINO3 time series, which is meant to represent the deviation of the SST from the monthly climatology, does not have a vanishing monthly mean climatology as one might expect. The monthly climatology of the perturbation model SST has an amplitude of 0.2 to 0.4 degree Celsius in many of the runs presented here (Tziperman et al, 1995). This perturbation climatology may be thought of as a sine wave-like time series with a 1 year frequency, superimposed on the interannual variability. If large enough, this superimposed seasonality may clearly affect the peak month of ENSO events, and thus affect ENSO's locking to the seasonal cycle. In order to prevent this, one may subtract the monthly climatology of the model SST time series, or use a 12 months running average before calculating the above mentioned histograms. However, we found that our conclusions concerning the model ENSO's locking to the seasonal cycle do not depend on the time series analysis procedure.

One of our main objectives in this work is to isolate the main mechanisms of interaction between the seasonal cycle and the interannual ENSO variability. The ZC model is a perturbation model for the deviations from the monthly climatology of the equatorial Pacific ocean and atmosphere. The fact that the background seasonal cycle is specified in the model rather than simulated may perhaps be considered a limitation of the model, yet it makes this model especially suitable for our purposes. Because the seasonal cycle in the various model fields is specified, we can conveniently adjust its amplitude as we shall do in the experiments described below.

There are five background fields that are specified as monthly climatologies in the model. Three of these are scalar fields: SST, oceanic upwelling and wind divergence; and two are horizontal vector fields: ocean currents and wind velocity. The monthly variability of these fields along the equator is plotted in Fig. 2 in the form of Hovmoller diagrams. All background fields are clearly seasonal at the equator, and the detailed seasonal structure of these fields will be further discussed below.

Let us consider now the way these background seasonal fields are incorporated into the model and examine the physical feedbacks controlled by each seasonal background field, following Zebiak and Cane (1987) and Zebiak (1986). Consider first the atmospheric heating which is assumed in the model to be dominated by moisture condensation and is divided into a contribution due to condensation of water evaporated locally at the ocean surface ($Q_s$) and another contribution due to the condensation of the larger scale humidity field due to the local wind convergence, $Q_1$. The total atmospheric heating is $Q = Q_s + Q_1$, where

$$Q_s = (\alpha T) \exp[\overline{T} - 30°C)/16.7°C] \tag{1}$$

$$Q_1 = \beta_*[M(\overline{c} + c) - M(\overline{c})] \tag{2}$$

$$c = -[(u_a)_x + (v_a)_y]. \tag{3}$$

In these equations, the monthly mean background convergence of the lower atmospheric winds is denoted by $\overline{c}(x, y, month)$, the perturbation convergence by $c(x, y, t)$, the zonal and meridional wind components are denoted by $u_a$ and $v_a$ correspondingly, and we define $M(x)$ such that $M(x) = x$ for $x > 0$ and $M(x) = 0$ otherwise. Clearly a larger mean SST ($\overline{T}$ in equation 1) will result in more heating $Q_s$ for the same perturbation SST ($T$ in the same equation). The dependence of the atmospheric heating on $\overline{T}$ is exponential, hence quite strong. The increase in climatological monthly equatorial eastern Pacific mean SST from about 23°C in September to about 26.5°C in March-April (Fig. 2a) corresponds to a 25% enhancement in the perturbation heating $Q_s$ for the same perturbation SST.

The seasonal variations in the mean wind convergence are potentially even more influential, and can cause the perturbation heating $Q_1$ to be turned on or off, depending on the size of the perturbation convergence relative to the background convergence. The seasonal variations in the mean convergence are a result of the seasonal motions of the ITCZ, which moves between about 10°N from July to September and just north of the equator earlier in the year (Philander, 1990). The southernmost location is reflected in Fig 2b as a negative divergence (equivalent, of course, to a positive convergence) signal along the equator in February to April. During these spring months, when the ITCZ is closer to the equator, there is a convergence of surface winds and therefore an upward air motion at the equator which heats the atmosphere through moisture condensation and latent heat release. The wind convergence enhances the coupling between sea surface temperature (SST) anomalies and atmospheric heating. Note that a positive equatorial SST anomaly, for example, creates atmospheric heating in the model due to the local condensation term $Q_s$ in (1). This heating induces convergent atmospheric motions and thus affect the wind stress driving the ocean currents. The coupling between SST and wind stress may be strongly enhanced for positive mean convergence, as in that case the direct atmospheric heating by local condensation is augmented by the heating due to the convergence of the large scale humidity field, through the term $Q_1$ in (2). This stronger coupling, in turn, enhances the coupled ocean-atmosphere instability process believed responsible for ENSO's onset. The months during which the ITCZ is closer to the equator are therefore clearly the months at which perturbation heating $Q_1$ will be maximized for a given perturbation convergence. On the other hand, when the ITCZ is further north of the equator, the mean divergence is positive, so that the perturbation wind divergence needs to be negative and larger in absolute value than the mean divergence in order to cause an atmospheric heating. In this state the SST and the atmospheric heating are largely decoupled from each other, weakening the coupled instability process.

The monthly mean atmospheric wind, $\overline{\mathbf{u}}_a$, appears in the expression for the perturbation wind stress $\tau$ applied to the ocean model momentum equations, as calculated from the atmospheric winds. Let the total wind velocity be $\mathbf{u}_a^* = (u_a^*, v_a^*) = \overline{\mathbf{u}}_a + \mathbf{u}_a$, then the perturbation wind stress is given by

$$\tau = (\tau^{(x)}, \tau^{(y)}) = \rho_{air} C_d \left( |\mathbf{u}_a^*| \mathbf{u}_a^* - |\overline{\mathbf{u}}_a| \overline{\mathbf{u}}_a \right) \quad (4)$$

Clearly, a larger mean wind velocity results in a larger perturbation wind stress, for a given size of the perturbation atmospheric wind $\mathbf{u}_a$. The monthly mean wind velocity $|\overline{\mathbf{u}}_a| = (\overline{u}^2 + \overline{v}^2)^{1/2}$ along the equator is shown in Fig. 2e.

The model equation for the sea surface temperature is

$$\frac{\partial T}{\partial t} = -\overline{\mathbf{v}}_1 \nabla T - \mathbf{v}_1 \nabla (\overline{T} + T) - \{M(\overline{w}_s + w_s) - M(\overline{w}_s)\} \overline{T}_z - M(\overline{w}_s + w_s) T_z - \alpha_s T. \quad (5)$$

Three background fields enter here: the SST ($\overline{T}$), ocean horizontal currents $\overline{\mathbf{v}}_1$, and upwelling $\overline{w}_s$. This equation is the only place where the background ocean currents and upwelling appear in the model equations, while as explained above, the background SST also appears in the heating parameterization used in the atmospheric model. The seasonal variation in the equatorial upwelling is shown in Fig. 2c, and shows some weakening of the upwelling signal in the central Pacific during March to May, and strengthening of the coastal upwelling in the eastern Pacific during August to September. The upwelling velocity serves to connect the thermocline thickness and thermocline-base temperature on one hand and the SST on the other. In this role, the upwelling thus connects ocean dynamics and coupled ocean-atmosphere thermodynamics. This has a potentially important effect on the coupled system stability, as emphasized by Battisti (1989). The seasonal variations of the zonal ocean currents is given in Fig. 2d, showing a pronounced seasonality over the entire equatorial Pacific.

The background SST enters the model physics in a third way in addition to the atmospheric heating parameterization (1) and the SST equation (5). Observations indicate that the monthly mean equatorial SST never appreciably exceeds 30 degrees Celsius. Accordingly, the model checks at every time step, after the SST is updated by the SST equation (5) that

$$\overline{T}(x, y, \text{month}) + T(x, y, t) \leq 30°. \quad (6)$$

If this constraint is not satisfied, the perturbation SST is reset to satisfy $\overline{T} + T = 30°$. This can be considered a simple parameterization of various cloud radiation effects which influence the air-sea heat flux, yet are not explicitly included in the model. Through this parameterization, the seasonal variations in the background SST can play a role in limiting the size of the perturbation SST.

Following Zebiak and Cane (1987), many of the effects of the seasonal background fields may be conveniently viewed as influencing the strength of coupling between the ocean and atmospheric components of the coupled model. For example, with a favorable (i.e. warm) background SST, a small SST perturbation results in a stronger atmospheric heating due to the local evaporation contribution $Q_s$ in (1). Thus a warmer background SST increases the coupling strength between the ocean and atmosphere. Similarly, a stronger background wind velocity amplifies the perturbation wind stress for a given perturbation wind velocity. As already mentioned above, the upwelling velocity strength can also enhance the coupled system instability by affecting the connection between ocean dynamics and the SST. We shall return later to this view of the effect of the background fields.

The importance of the seasonal background fields in forcing the irregularity of the ZC model ENSO events was demonstrated in Tziperman et al. (1995). It was also shown there that when all background fields are set to their annual mean, the background state is too stable and cannot support model ENSO events, therefore resulting in a zero model solution. The question that arises is therefore: which of the background fields' seasonality is the most crucial to the existence of model ENSO events, to their chaotic behavior, and to their locking to the seasonal cycle? This question is answered in the following sections.

# 3    First order seasonal effect: the wind divergence

We consider now a set of experiments in each of which only one of the five background fields is set to its full monthly variability, while all other four are set to their annual average, and thus vary in space but not in time. Before describing the results, a comment is due concerning the self consistency of our approach. Clearly all of the climatological background fields specified in the model are mutually dependent, and it does not make physical sense to make some of them seasonal and the others not. As an especially obvious example we note that if the horizontal ocean currents are made seasonal but the upwelling is not, then the background ocean velocity field does not satisfy the continuity equation. While we are perfectly aware of this difficulty, we still feel that important lessons can be learned using the artificial separation between the various background fields. Such a separation is technically possible only because the ZC model is a perturbation model in which the monthly climatology is specified rather than calculated. We invite the reader to bear with us and allow us to demonstrate that the procedure we use, while somewhat artificial, is indeed useful.

The results of all experiments discussed in this work are summarized in Table 1. In the first experiment (run SST in Table 1), only the background SST is seasonal, while the other background fields are set to their annual average. The resulting NINO3 time series (Fig. 3a) shows that ENSO events are weak, typically 0.8 to 1 degree amplitude. The events are locked to July (Fig. 3c) rather than Oct-Dec in the standard model solution. The time series is chaotic and events occur every 2.5-5.5 years (Fig. 3d). Overall, this solution differs enough from the standard solution (Fig. 1) to indicate that the seasonality in the SST is not the main seasonal forcing of model ENSO events.

In the next experiment (run DIV in Table 1), only the background wind divergence is seasonal. The model ENSO events are much stronger now, with a typical amplitude of about 3 degrees (Fig. 4a). The time series is chaotic and is characterized by a broad spectrum (Fig. 4b). There is a broad locking of the model ENSO events to May-Nov (Fig. 4c) and most events occur every about 2 to 6 years (Fig. 4d). Overall, this solution is quite similar to the standard model solution of Fig. 1.

In the experiments UPWEL and WIND of Table 1, in which only the ocean upwelling and the atmospheric wind are set to be seasonal, correspondingly, with other fields set to their annual average, the background state is too stable, ENSO events cannot develop and the model solution rapidly converges to zero.

The final experiment in this series (run UV in Table 1) sets the horizontal ocean currents to be seasonal. In this case only very weak ENSO events can develop, with amplitude of about 0.3 degree (Fig. 5a). The time series is weakly chaotic (see sharp peaks in power spectrum, Fig. 5b) and model ENSO events are locked to September (Fig. 5c). These events occur every 3 or 4 years. Clearly the seasonality in the ocean currents does not provide the background state required for the development of more realistic ENSO events.

This set of experiments indicates that the main background seasonal effect in the ZC model seems to be due to the wind divergence field. Physically, this tells us that the atmospheric heating due to the condensation of the (implied) large scale humidity field is a dominant mechanism by which the background state is made unstable during certain periods of the calendar year. Our experiments may also be viewed as an attempt to find which background field provides the necessary instability to support ENSO events. This seasonal instability of the background state, resulting from the seasonal motion of the ITCZ, allows the development of ENSO events, and is also reflected in ENSO predictability studies (Blumenthal, 1991; Xue et al, 1994, Goswami and Shukla, 1991; Battisti and Sarachik, 1995; Latif et al, 1994; Chen et al, 1995).

While the seasonal cycle was often ignored in delayed-oscillator type theories, it is interesting to note that Cane et al. (1990) and Munnich et al. (1991) have incorporated it as a relatively small (25%) modulation of the coupling strength between the ocean and the atmosphere. However, we have emphasized here that both the annual-mean background state, and the background state during several months during the year, are stable and cannot support ENSO events. Delayed oscillator models typically use a coupling coefficient whose value determines the stability of the background state. Only above some critical value of this coupling coefficient does the coupled instability occur, enabling self sustained ENSO events. Our results concerning the stability of the background state in the ZC model imply that if we were to use the coupling strength concept in a delayed oscillator model, the seasonal changes in the coupling coefficient would have to bring the value of the coefficient to below its critical value for certain times during the year. During these times the development of the coupled instability would be prohibited, as is the case with the ZC model.

In order to further examine the seasonal variability of the background stability, we have run a series of twelve perpetual month experiments in each of which the background fields were set to a given monthly climatology. We have calculated the averaged event amplitude for each of these runs, and have plotted it against the month used as background climatology (Fig. 6). We tend to think of the amplitude of ENSO events for this series of experiments as representing a rough measure of the stability of the background state represented by each of the monthly climatologies. According to this view the background provided by the more unstable months can support larger amplitude events while the stable months cannot support self sustained ENSO variability or only allow the development of weak events (a more rigorous stability analysis of the ZC model was carried out in Blumenthal, 1991 and Xue et al, 1994). As seen in Fig. 6, the most unstable months seem to be May to August. February and September are too stable to support ENSO events, and October, November and January allow relatively weak events. Some of the seasonal variability of the basic state stability is due to the seasonal variability of the wind divergence, due to the ITCZ motion. But the seasonal variations in the ITCZ cannot account for the full behavior seen in Fig. 6, as the wind divergence mostly destabilizes the background state during March to April when the equatorial divergence is negative (Fig. 2b), while the fully seasonal background model state seems most unstable from March to August. We conclude that the background stability is set by additional factors to the background wind divergence.

While it is quite clear from this first set of experiments that the seasonality of the background wind convergence field is the dominant background seasonal effect in the model, it is certainly not the only such seasonal effect. The locking to the seasonal cycle in run DIV (Fig. 4c) is unlike that of the standard model solution, and it seems that there must be additional seasonal forcings that affect and shape the standard model solution. The above discussion of the factors determining the background model stability also indicated that there must be additional seasonal effects at work. The next section attempts to isolate these additional seasonal effects.

## 4  Second order seasonal effects: the SST

We now consider a second set of experiments in each of which two background fields are set to be fully seasonal: the background wind divergence and a second background field. In examining the results of these runs, we will be looking in particular for a second seasonal background field that produces the locking of model ENSO events to the end of the calendar year which was missing when only the wind divergence was seasonal in run DIV of Table 1. This present set of experiments could be seen as an effort to define the least amount of background seasonality that can reproduce the main features of the standard model run.

The first of this series of experiments (Fig. 7, run DIV+UV in Table 1) sets both the wind divergence and the horizontal ocean currents to their monthly climatology while other fields are set to their annual averages. The result is close to that obtained in run DIV, except that locking of ENSO events to May-November is a bit more pronounced now. This partial locking of model ENSO events to the seasonal cycle is still not particularly close to the standard model solution.

In the second experiment (run DIV+SST), the background wind divergence and background SST are the only seasonal fields. The results are very close to the standard model solution. There is strong locking to October and November, and the NINO3 time series has an irregular character as in the standard solution. On the other hand, model ENSO events in this run tend to occur mostly every 3 years (Fig. 8d) rather than mostly every 4 years as in the standard model run, and the averaged event amplitude is somewhat too large (Table 1). Apart from these last two points, which will be shown later to be due to the missing seasonality in the upwelling velocity, the solution is fairly reasonable. This, in fact, is the most satisfactory case of this set of experiments. The shift of the model frequency from 4 to 3 years is an example of the sensitivity of this frequency in the presence of a seasonal background. This sensitivity may be explained as a tendency of the model to shift between different nonlinear resonances with the annual cycle, which is the essence of the mechanism believed responsible for ENSO's irregularity in the ZC model (Tziperman et al, 1995).

Experiment DIV+UPWEL (Fig. 9) sets the ocean upwelling to be the second seasonal field in addition to the wind divergence. The results are characterized by a weak locking of ENSO events to March-August, and are thus not satisfactory.

In the last experiment in this series (run DIV+WIND of Table 1, Fig. 10), the background wind divergence and background wind velocity are seasonal. The ENSO events tend to occur in August to November as well as in January and February. Again not sufficiently close to the standard case.

We conclude from this set of experiments that two most important seasonal influences on the interannual model variability seem to be the climatological seasonal cycle in (i) the wind divergence and (ii) the SST. While the effect of the seasonality in the background divergence field is fairly clear and enters the model equations only in the atmospheric heating formulation as discussed above, the way seasonality in the background SST influences the interannual variability is not as obvious. The background SST enters the model equations in the atmospheric heating formulation, in the SST equation in which the climatological SST is advected by the perturbation currents, and in the 30° limit imposed on the total SST. In order to isolate the main SST effect, we carried out three additional model experiments. In all three experiments the wind divergence was fully seasonal. In addition, the first experiment (DIV+SST-heat in Table 1), incorporates the background SST seasonality only in the heating parameterization $Q_s$ in (1), while in the SST equation (5) and in the constraint (6), the background SST was set to its annual average. In the second experiment (DIV+SST-eqn in Table 1), the SST was set to be fully seasonal only in the SST equation (5). Finally, in the third experiment (DIV+SST-lim in Table 1), the background SST was fully seasonal only when used in the 30° limit on the total SST (6).

Fig. 8e-g show the histograms with the information on the locking of ENSO events to the calendar year in these three runs. The experiment DIV+SST-heat in which the background SST enters in the heating parameterization resulted in locking to June to September rather than to the end of the year (Table 1 and Fig. 8e). The other two experiments in which the background SST is set to be seasonal in the SST equation (Fig. 8f) and in the 30° limit (Fig. 8g) are closer to the standard run in terms of the locking of model ENSO events to the annual cycle. It is difficult to judge which of the two produces a closer solution to the standard run. It seems that both of the seasonal background SST effects in these two experiments, i.e. the advection of background seasonal SST by the perturbation currents and upwelling, and the effect of the background SST seasonality through the 30° limit, are important.

The results of a complementary set of experiments (Table 1, experiments NO:SST-heat, NO:SST-eqn and NO:SST-lim) intended to examine which is the dominant seasonal SST feedback are shown in Fig. 8h-j. In experiment NO:SST-heat, all background fields are fully seasonal, except for the SST background in the atmospheric heating parameterization that is set to its annual mean (Fig. 8h). Similarly in experiment NO:SST-eqn, only the SST background used in the SST equation is set to its annual mean (Fig. 8i). Run NO:SST-lim sets the SST to annual mean only in the 30° limit while all other background fields are fully seasonal (Fig. 8j). Overall, these experiments support the conclusions derived from Fig. 8e-g. When only the SST in the atmospheric heating is set to annual mean (Fig. 8h), the model ENSO locking is to October to December, as in the standard solution, indicating that the seasonal variation of the SST in the heating parameterization may not play an important role. We note, however, that the locking did shift from being mostly in October to being mostly in December, indicating that the seasonality of the SST in the heating parameterization may play a non-negligible role after all. The shift of locking when the SST seasonality is removed in the other two feedbacks (SST equation, Fig. 8i, and 30° limit, Fig. 8j) is somewhat more significant, indicating again that these two feedbacks may be more dominant in the model.

The above results for the particular SST feedbacks that are most dominant in setting the model ENSO characteristics are probably the most model-dependent of the results presented in this work and should be viewed with some degree of skepticism. In any case, it seems that the two SST background seasonal effects found dominant are not strongly related to the atmospheric heating parameterization through which the background seasonality in the wind divergence interacts with the interannual variability. It seems that the precise mechanism that leads to partial locking of model ENSO events to the annual cycle is fairly complex and involves several different effects.

# 5  Seasonal effects due to the ocean upwelling

We noted in the last section that in the run where both the wind divergence and the SST were seasonal, the solution was quite close to the standard case, apart from the too large event amplitude, and the 3 year ENSO frequency instead of the 4 year frequency in the standard run. In order to isolate the reason for these deviations from the standard model solution (and from the observed ENSO characteristics), we have run a third series of experiments in which only one background field was set to annual mean, while the others were fully seasonal (run names beginning with "NO:" in Table 1). The results basically confirmed the findings of the previous two sections concerning the main seasonal effects, and we will not show the detailed analysis of these runs. A summary analysis of these runs is given in Table 1. One important finding of these experiments is that when the ocean upwelling velocity is the only field set to its annual average, then the event amplitude rises significantly, and the frequency of events shifts to 3 years (run NO:UPWEL in Table 1). This is consistent with Battisti's (1989) proposed influence of the upwelling velocity on the coupled instability mechanism as explained above. This result also seems to suggest that the remaining unsatisfying characteristics of run (DIV+SST) with only the wind divergence and SST being seasonal are probably due to the lack of seasonality in the background upwelling. An experiment in which the wind divergence, SST and ocean upwelling are all seasonal (DIV+SST+UPWEL in Table 1) was indeed characterized by a dominant four year frequency, and by smaller amplitude events than in the absence of the upwelling seasonality.

While the upwelling seems to have an important role in our experiments, it does not play as important a role as in the analysis of Battisti (1989), but only a secondary role to that of the wind divergence and SST. It is interesting that in Battisti's model, which is closely related to the ZC model used here, the annual mean background does support self sustained oscillations, unlike Hirst's (1986) analysis and our results here. The seasonality of the background wind divergence played a major role in our analysis mostly because it created times during the year when the coupled instability could not exist, and appears to be the main cause of the inability of the annual mean state to support the developments of ENSO events. This may be related to the seeming difference between the major seasonal effects in the two models. Other differences between the two models were thoroughly investigated by Mantura and Battisti (1995).

# 6 Conclusions

That the seasonal cycle in the equatorial Pacific affects the interannual ENSO variability is fairly obvious from the partial locking of ENSO events to the seasonal cycle (Rasmusson and Carpenter, 1982), as well as from previous works suggesting that the seasonal cycle forces ENSO's irregularity (Tziperman et al., 1994, Jin et al., 1994, Tziperman et al. 1995, Chang et al., 1994, Chang et al, 1995). We attempted here to identify the physical mechanisms by which the seasonality of the background state of the equatorial Pacific influences the interannual variability in a particular coupled ocean atmosphere ENSO model – that of Zebiak and Cane (1987).

First, we determined which of the seasonally varying climatological fields specified as a background in the ZC model is most crucial to the model ENSO's dynamics. The answer seems to be that the seasonal changes in the background wind divergence are the main (first order) seasonal effect. Second order seasonal effects due to the seasonality of the background SST were also found to be essential in setting the precise locking of model ENSO events to the calendar year. Finally, the seasonality of the background climatological upwelling velocity was also found to have a significant effect on the frequency and amplitude of the model ENSO events.

The mechanisms by which the wind divergence and SST interact with the interannual variability were identified in detail. We emphasized that the climatological background state specified in the model is stable and cannot support the coupled instability and the development of model ENSO events at certain periods during the year. The annual mean background state is also too stable to support self sustained ENSO events. This means that if we were to use a "coupling coefficient" within an idealized delayed oscillator model (Cane et al, 1990; Munich et al, 1991), the seasonal changes in the coupling coefficient would have to bring the value of the coefficient to below its critical value at certain times during the year. During these times, when the coupling strength is below its critical value, the development of the coupled instability would be prohibited, as is the case with the ZC model or in the stability analysis of Hirst (1986). This represents a significant modification of the delay oscillator concept developed in previous studies.

While we feel we were able to improve our understanding of the mechanisms of seasonal-ENSO interaction beyond the general understanding provided by the recent works on the importance of the seasonal cycle to ENSO's irregularity, there are still quite a few open questions. It would be useful, for example, to obtain a better understanding of the spatial structure of the seasonal-ENSO interaction mechanisms discussed here. In addition, better physical insight could be gained if a complementary approach to the seasonal-ENSO interaction mechanisms could be developed from the point of view of equatorial ocean waves.

Finally, our analysis and results centered on a specific model. We found that even in this simplified model the mechanisms of seasonal-interannual interaction are not simple to isolate. We can expect this interaction to be even more complex in the actual Pacific ocean and atmosphere. We hope that the analysis presented here will help to guide studies using more complex models as well as data analysis efforts addressing the interaction between the seasonal cycle and the interannual variability in the equatorial Pacific.

**Acknowledgments:**

We thank Larry Rosen for his help, and Ilya Rivin for useful discussions. This work is partially supported by a grant from NOAA through the Consortium for Climate Research, UCSIOPO-10775411D/ NA47GP0188. SZ and MC are also partially supported by NSF grant ATM92-24915.

# 7 Bibliography

Battisti, D. S., 1989: The dynamics and thermodynamics of a warming event in a coupled tropical atmosphere/ ocean model. J. Atmos. Sci., 45, 2889-919.

Battisti, D. S, and Hirst, A. C., 1989: Interannual variability in the tropical atmosphere/ ocean system: influence of the basic state, ocean geometry and nonlinearity. *J. Atmos. Sci.*, **46**, 1687-1712

Battisti, D.S. and E.S. Sarachik, 1995: Understanding and predicting ENSO. Reviews of Geophysics, Supplement, Vol. 33, 1367-1376.

Blumenthal, M. b., 1991: Predictability of a coupled ocean-atmosphere model. J. Climate 4, 766-84.

Cane, M. A., M. Munnich and S. E. Zebiak, 1990: A study of self-excited oscillations of the tropical ocean-atmosphere system. Part I: linear analysis. *J. Atmos. Sci.*, **47**, 1562-1577.

Chang P., B. Wang., T. Li, L. Ji., 1994: Interactions between the seasonal cycle and the southern oscillation — frequency entrainment and chaos in a coupled ocean-atmosphere model. Geophys Res Let, 21, 2817-2820.

Chang P., Link Ji, Bin Wang and Tim Li, 1995: On the interactions between the seasonal cycle and El NiñoSouthern Oscillation in an intermediate coupled ocean-atmosphere model. *J. Atmos. Sci.* in press.

Chen, D., S.E.Zebiak, A.J. Busalacchi and M.A Cane, 1995: An Improved Procedure for El Nino Forecasting: Implications for predictability. to appear in Science.

Goswami, B.N., and J. Shukla, 1991: Predictability of a coupled ocean-atmosphere model. J. Climate, 4, 107-115.

Graham N. E. and W. B. White, 1988: The El Niño cycle: A natural oscillator of the Pacific ocean-atmosphere system. *Science*, **240**, 1293-1302.

Hirst, A. C., 1986: Unstable and damped equatorial modes in simple coupled ocean-atmosphere models. *J. Atmos. Sci.*, **43**, 606-630.

Jin, F-F, D. Neelin and M. Ghil, 1994: ENSO on the devil's staircase. *Science*, **264**, 70-72.

Latif, M. T.P. Barnett, M.A. Cane, M. Flugel, N.E. Graham, H. von Storch, J.-S. Xu, and S.E Zebiak, 1994: A review of Enso prediction studies. Climate Dynamics, 9, 167-179.

Mantura N., J. and D. S. Battisti, 1995: Aperiodic variability in the Zebiak-Cane coupled ocean-atmosphere model: air-sea interactions in the western equatorial Pacific. J. Climate, Submitted.

Munnich, M., M. A. Cane and S. E. Zebiak, 1991: A study of self-excited oscillations of the tropical ocean-atmosphere system. Part II: nonlinear cases. *J. Atmos. Sci.*, **48**, 1238-1248.

Neelin, J. D., 1991: the slow sea surface temperature mode and the fast-wave limit: analytic theory for tropical interannual oscillations and experiments in a hybrid coupled model. J. Atmos. Sci. 48, 584-606.

Philander, S. G., 1983: El Niño Southern Oscillation phenomena. *Nature*, **302**, 295-301.

Philander, S. G., 1990: *El Niño, La Niña, and the Southern Oscillations*, Academic Press, San Diego, California.

Philander, S. G., T. Yamagata and R. C. Pacanowski, 1984: Unstable air-sea interactions in the tropics. *J. Atmos. Sci.*, **41**, 604-613.

Rasmusson, E., and T. Carpenter, 1982: variations in the tropical sea surface temperature and surface wind fields associated the the Southern Oscillations/ El Niño. *Mon. Wea. Rev.*, **110**, 354-384.

Suarez, M. J., and P. S. Schopf, 1988: A delayed action oscillator for ENSO. *J. Atmos. Sci.*, **45**, 3283-3287.

Tziperman E., L. Stone, M. Cane and H. Jarosh, 1994: El Niño chaos: Overlapping of resonances between the seasonal cycle and the Pacific ocean-atmosphere oscillator. *Science*, **264**, 72-74.

Tziperman, E., M. A. Cane and S. Zebiak, 1995: Irregularity and locking to the seasonal cycle in an ENSO prediction model as explained by the quasi-periodicity route to chaos. *Journal of the Atmospheric Sciences*, **52**, 293-306.

Xue, Y., M. A. Cane, S. E. Zebiak, and M. B. Blumenthal, 1994: On the prediction of ENSO: a study with a low-order Markov model. Tellus, 46A, 512-28.

Zebiak, S. E., 1986: Atmospheric convergence feedback in a simple atmospheric model for El Niño *Monthly Weather Review*, 114, 1263-71.

Zebiak, S. E. and M. Cane, 1987: A model El Niño-Southern Oscillation. *Mon. Weather Rev.*, **115**, 2262

# 8 Figure Captions

Figure 1: An analysis of a NINO3 time series from a 1024 year standard model run: (a) a portion of the NINO3 time series, (b) power spectrum, (c) A histogram of the number of ENSO events (vertical axis) per month of the calendar year (horizontal axis). (d) A histogram of the distribution of separation between model ENSO events. Horizontal axis: separation between events in years (in a 3 months resolution); vertical axis: number of times a given separation is seen in the time series.

Figure 2: The monthly variability along the equator of the climatological fields specified as background in the ZC model. (a) SST ($°C$), (b) wind divergence ($10^{-6}s^{-1}$), (c) upwelling ($10^{-5}m/s$), (d) zonal ocean velocity ($.1m/s$), (e) zonal wind velocity, $(\overline{u}^2 + \overline{v}^2)^{1/2}$ ($m/s$).

Figure 3: First order seasonal effects: Same as Fig. 3 for run SST of Table 1.

Figure 4: First order seasonal effects: Same as Fig. 3 for run DIV of Table 1.

Figure 5: First order seasonal effects: Same as Fig. 3 for run DIV of Table 1.

Figure 6: Averaged model ENSO amplitude for twelve perpetual month experiments.

Figure 7: Second order seasonal effects: Same as Fig. 3 for run DIV+UV of Table 1.

Figure 8: (a)-(d) Second order seasonal effects: Same as Fig. 3 for run DIV+SST of Table 1. (e) Monthly histogram for run DIV+SST-heat. (f) Monthly histogram for run DIV+SST-equation. (g) Monthly histogram for run DIV+SST-limit. (h) Monthly histogram for run NO:DIV+SST-heat. (i) Monthly histogram for run NO:DIV+SST-equation. (j) Monthly histogram for run NO:DIV+SST-limit.

Figure 9: Second order seasonal effects: Same as Fig. 3 for run DIV+UPWEL of Table 1.

Figure 10: Second order seasonal effects: Same as Fig. 3 for run DIV+WIND of Table 1.

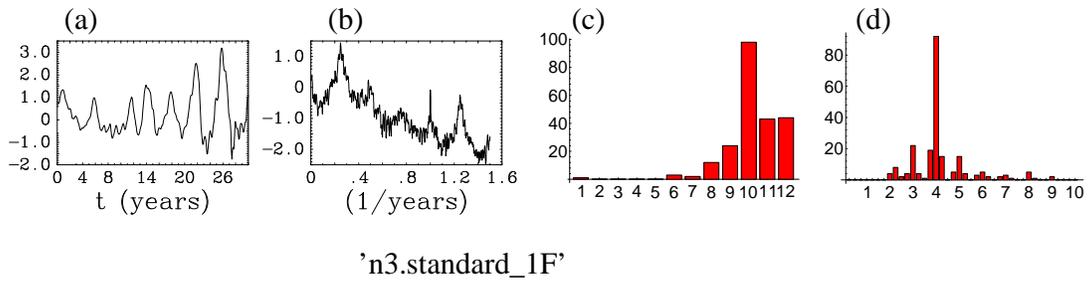

'n3.standard_1F'

Figure 1

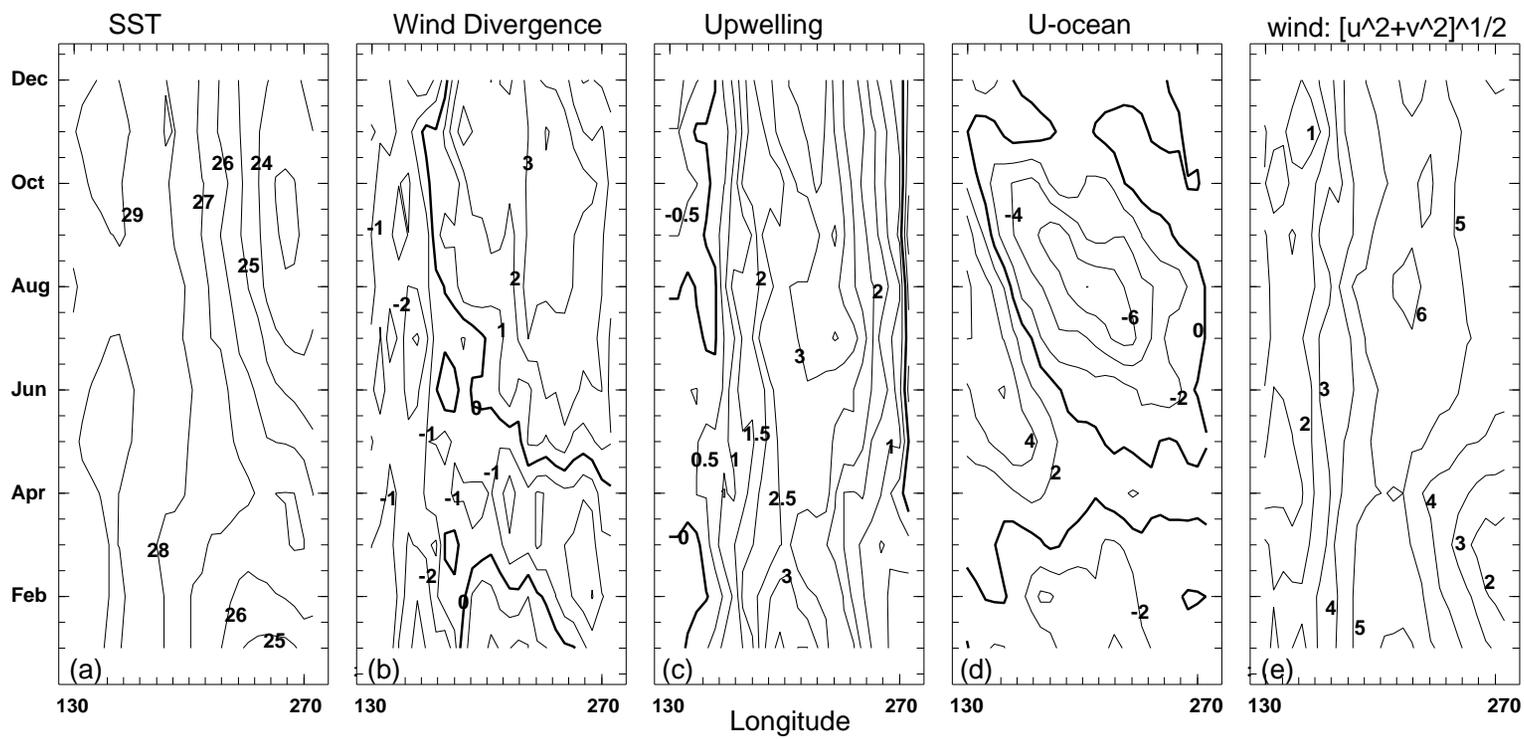

Figure 2

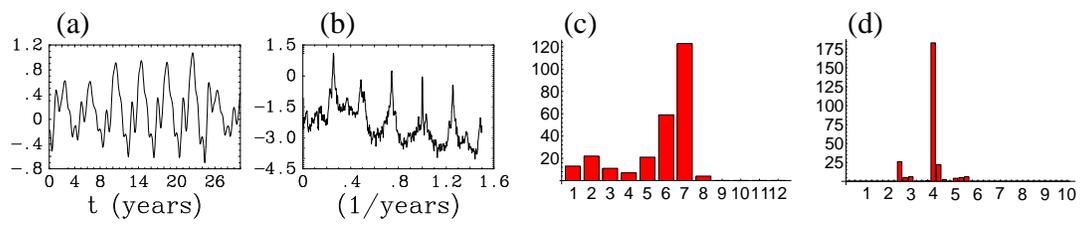

'n3.standard_43aF'

Figure 3

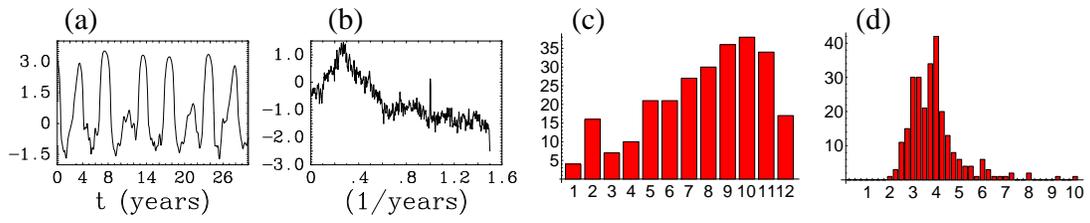

'n3.standard_43bF'

Figure 4

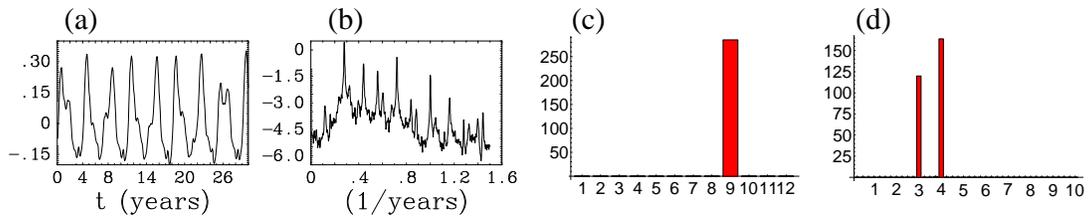

'n3.standard_43eF'

Figure 5

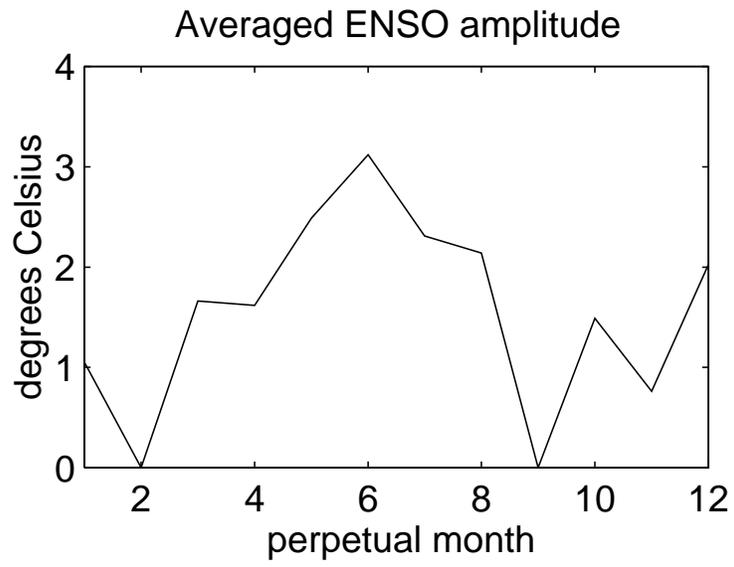

Figure 6

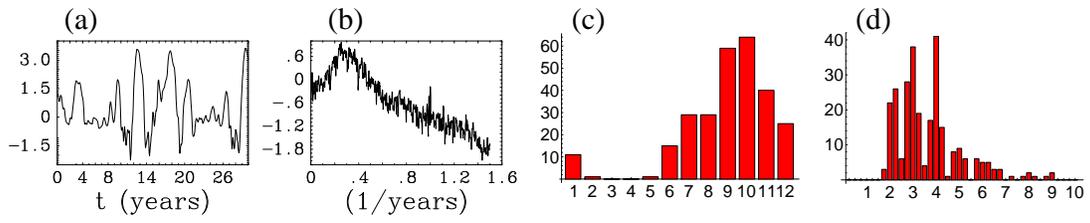

'n3.standard_50aF'

Figure 7

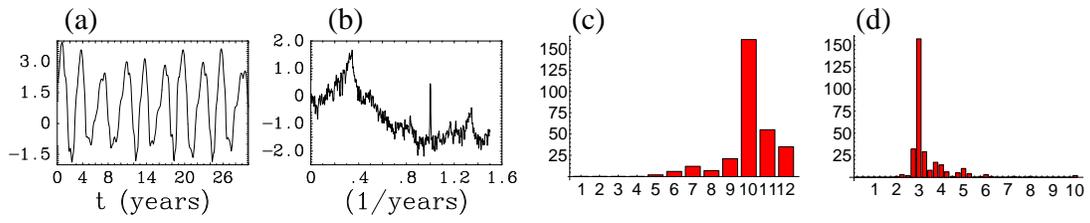

'n3.standard_50bF'

Figure 8a-d

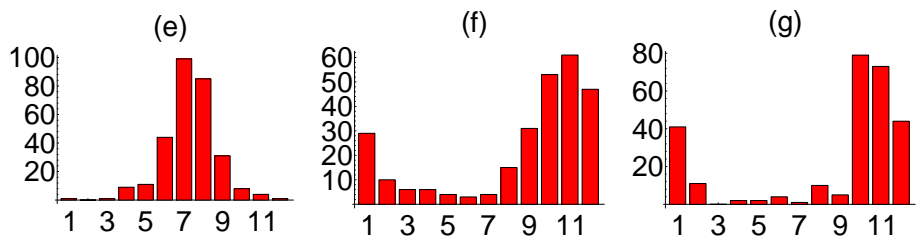

Figure 8e-f

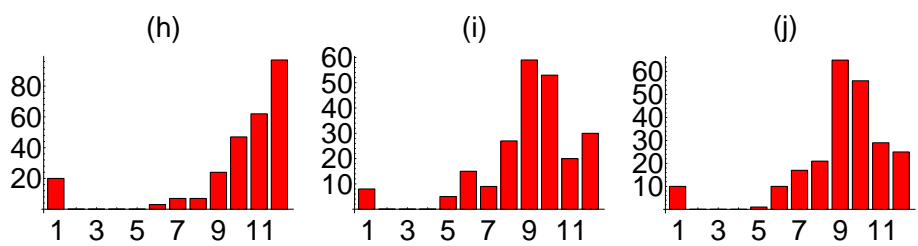

Figure 8h-j

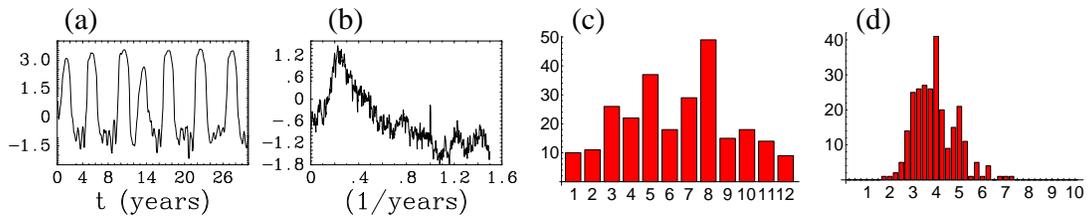

'n3.standard_50cF'

Figure 9

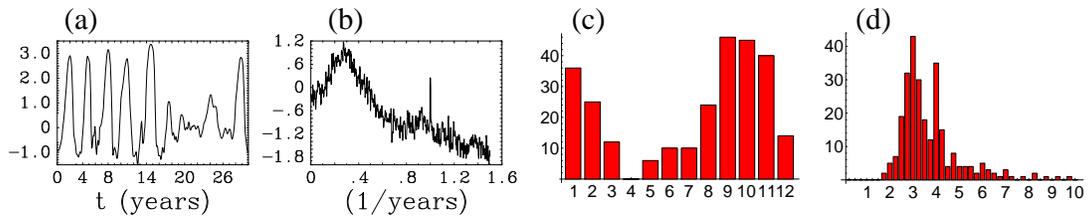

'n3.standard_50dF'

Figure 10

| Run | Seasonal Fields | Averaged Event (°C) | Locking to Seasonal Cycle | Fig. |
| --- | --- | --- | --- | --- |
| Standard | all | 2.1 | Oct-Dec | 1 |
| Annual | none | 0.0 | - | - |
| SST | SST | 0.8 | Jun-Jul | 3 |
| DIV | wind divergence | 2.8 | May-Dec | 4 |
| UPWEL | ocean upwelling | 0.0 | - | - |
| WIND | wind velocity | 0.0 | - | - |
| UV | ocean currents | 0.3 | Sep | 5 |
| perpet | (perpetual month background) | (See Fig. 6) | - | 6 |
| DIV+UV | div & currents | 2.4 | Jul-Dec | 7 |
| DIV+SST | div & SST | 3.1 | Oct-Nov | 8a-d |
| DIV+SST-heat | div&SST heating | 2.8 | Jun-Sep | 8e |
| DIV+SST-eqn | div&SST equation | 2.6 | Sep-Dec | 8f |
| DIV+SST-lim | div&SST 30°limit | 2.7 | Oct-Jan | 8g |
| DIV+UPWEL | div & upwelling | 2.8 | (Mar-Aug) | 9 |
| DIV+WIND | div & wind veloc. | 2.4 | Aug-Nov&Jan-Feb | 10 |
| DIV+SST+UPWEL | div, SST &upwelling | 2.6 | Oct-Dec | - |
| NO:UV | all but currents | 2.6 | Sep-Dec | - |
| NO:DIV | all but wind div | 0.0 | - | - |
| NO:SST | all but SST | 1.9 | Jun-Jan | - |
| NO:SST-heat | all but SST heating | 1.8 | Oct-Dec | 8h |
| NO:SST-eqn | all but SST equation | 1.9 | Aug-Dec | 8i |
| NO:SST-lim | all but SST 30°limit | 1.9 | Sep-Oct | 8j |
| NO:UPWEL | all but upwelling | 3.5 | Oct | - |
| NO:WIND | all but wind veloc. | 2.0 | Sep-Nov | - |

Table 1: Summary of model runs used in this study.